\documentclass[preprint,12pt]{elsarticle}

\usepackage[utf8]{inputenc}    
\usepackage[T1]{fontenc}       
\usepackage{lmodern}           
\usepackage{textcomp}          
\usepackage{amssymb, amsmath}  
\usepackage{microtype}         
\usepackage{pifont}            

\usepackage{booktabs}          
\usepackage{array}             
\usepackage{tabularx}          
\newcolumntype{Y}{>{\raggedright\arraybackslash}X}

\usepackage{graphicx}
\usepackage{caption}
\usepackage{subcaption}
\usepackage{float}

\usepackage{siunitx}
\usepackage{xcolor}
\usepackage{orcidlink}
\usepackage{hyperref}
\hypersetup{
  colorlinks=true,
  linkcolor=blue!60!black,
  citecolor=blue!60!black,
  urlcolor=blue!60!black,
  pdftitle={Lightweight Session-Key Rekeying Framework for Secure IoT–Edge Communication},
  pdfauthor={Haranath Rakshit, Subhasis Banerjee}
}



\usepackage[margin=2.5cm]{geometry}   

\begin{document}
\begin{frontmatter}

\title{Lightweight Session-Key Rekeying Framework for Secure IoT–Edge Communication}

\author[aff1]{Haranath Rakshit\orcidlink{0009-0008-1571-4440}\corref{cor}\fnref{first}}

\author[aff1]{ Rajkumar Bhandari\orcidlink{0009-0009-3912-4389}\fnref{eq1}}

\author[aff1]{Subhasis Banerjee\orcidlink{0009-0007-1154-7289}\fnref{eq2}}

\affiliation[aff1]{
  organization={Department of Computer and System Sciences, Siksha-Bhavana, Visva-Bharati},
  addressline={Santiniketan},
  city={Bolpur},
  postcode={731235},
  state={West Bengal},
  country={India}
}

\cortext[cor]{Corresponding Author. \href{mailto:haranathrakshit@gmail.com}{haranathrakshit@gmail.com}}
\fntext[first]{First Author.}

\fntext[eq1]{Contributing Author. \href{mailto:rajkumarbhandari.rs.css@visva-bharati.ac.in}{rajkumarbhandari.rs.css@visva-bharati.ac.in}}

\fntext[eq2]{Contributing Author. \href{mailto:subhasis.banerjee@visva-bharati.ac.in}{subhasis.banerjee@visva-bharati.ac.in}}

\begin{graphicalabstract}
\centering
\includegraphics[width=1\textwidth]{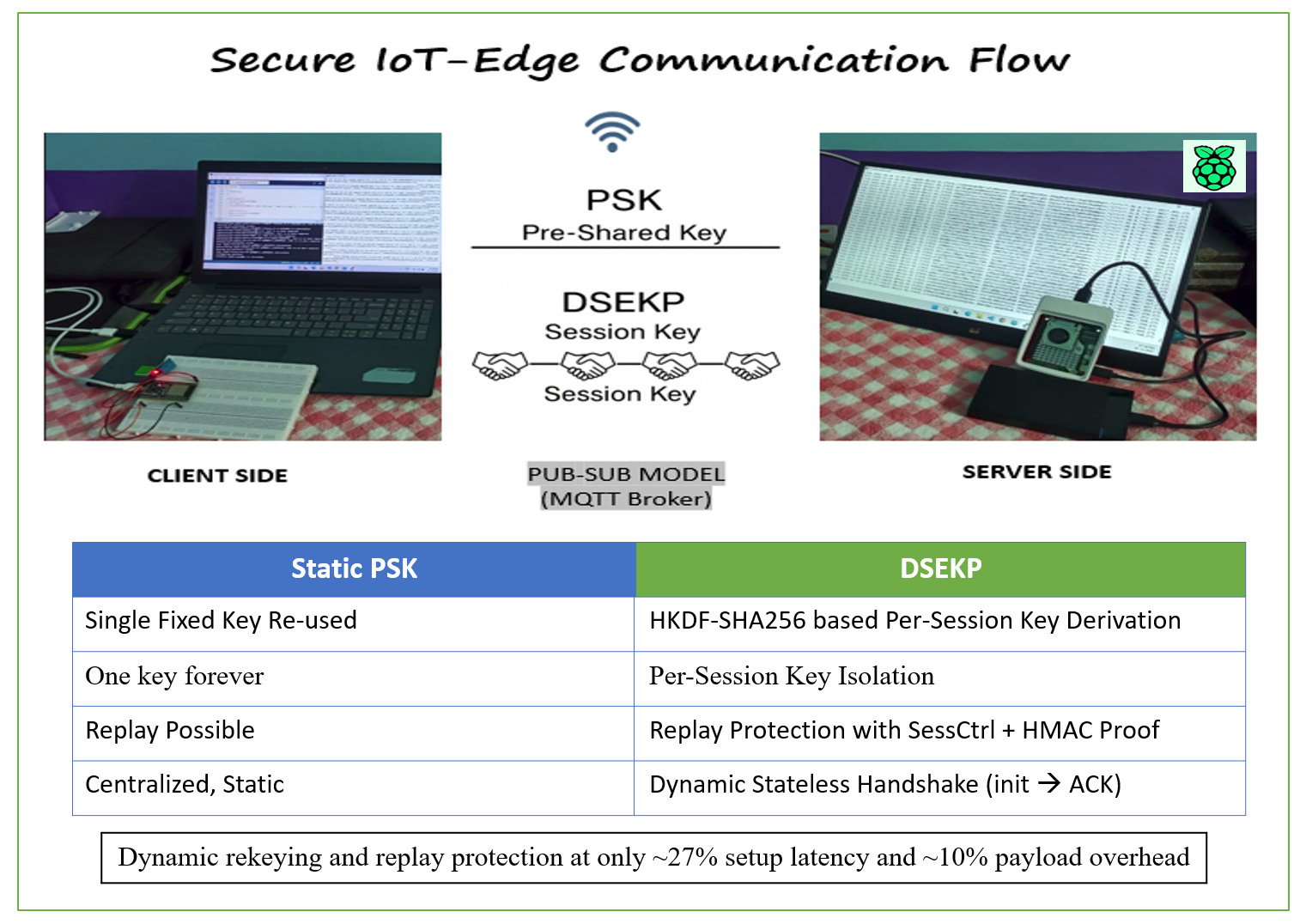}
\end{graphicalabstract}

\begin{highlights}
\item HKDF-derived session keys strengthen lightweight IoT–Edge security.
\item AES–GCM provides per-session confidentiality and integrity.
\item Achieves authenticated key refresh without public-key operations (single init–ack handshake).
\item Adds only $\approx$27\% latency and 10\% payload overhead compared with static PSK.
\item Validated experimentally on an ESP32–Raspberry~Pi~5 testbed over 6{,}500 encrypted packets.
\end{highlights}

\begin{abstract}
The proliferation of Internet of Things (IoT) networks demands security mechanisms that protect constrained devices without the computational cost of public-key cryptography. Conventional Pre-Shared Key (PSK) encryption, while efficient, remains vulnerable due to static key reuse, replay attacks, and the lack of key freshness. This paper presents the \textit{Dynamic Session Enhanced Key Protocol (DSEKP)}—a lightweight session-key rekeying framework that derives per-session AES--GCM keys using the HMAC-based Key Derivation Function (HKDF--SHA256) and authenticates session establishment through an HMAC proof in a single init--ack exchange. DSEKP was implemented on an ESP32 IoT sensor node and a Raspberry~Pi~5 edge server communicating through a Mosquitto MQTT broker, and benchmarked against a static PSK baseline over more than 6{,}500 encrypted packets per configuration. The results demonstrate nearly identical throughput and reliability, with minimal runtime impact---$\approx$27\% one-time session-establishment latency and $\approx$10\% per-packet payload overhead---while delivering per-session key isolation (assuming the long-term device secret remains uncompromised) and built-in replay protection. The PSK baseline and DSEKP datasets are publicly released to ensure reproducibility of experimental evaluation. These findings confirm that dynamic symmetric rekeying can substantially strengthen IoT--Edge links with minimal computational and bandwidth cost, offering a practical migration path from static PSK to session-aware, scalable, and reproducible IoT security.
\end{abstract}


\end{frontmatter}


\section{Introduction}
\label{sec:intro}

The rapid evolution of the Internet of Things (IoT) has transformed everyday environments into cyber–physical ecosystems that enable automation, data analytics, and intelligent decision-making across domains such as healthcare, smart cities, and industrial control. However, this large-scale interconnectivity introduces major challenges in maintaining the confidentiality, integrity, and authenticity of communications among resource-constrained edge devices.

IoT nodes such as microcontrollers and sensors operate with limited memory, processing power, and energy resources, rendering traditional public-key cryptographic protocols—such as Transport Layer Security (TLS) or Datagram TLS (DTLS) non-efficient~\cite{han2015semantic} for large-scale deployment. Consequently, symmetric encryption based on the Pre–Shared Key (PSK) model remains a prevalent choice for securing lightweight protocols like MQTT and CoAP ~\cite{shyam2025survey,bideh2020energy} due to its simplicity, low computational cost, and deterministic timing behavior.

Despite its efficiency, static PSK encryption exhibits fundamental weaknesses. The reuse of a single long-term key across all sessions compromises both past and future communications once the key is exposed~\cite{kuo2006comparison}. Moreover, static PSK reuses a single long-term key, exposing all past and future encrypted traffic if the key is compromised, and lacks replay protection when no freshness mechanisms are applied. The operational challenges of securely distributing, rotating, and updating PSKs further hinder scalability in dynamic IoT networks~\cite{michaels2018high}.

To overcome these limitations, this study introduces the \textbf{Dynamic Session Enhanced Key Protocol (DSEKP)}—a lightweight, symmetric-only enhancement to the PSK framework that employs the HMAC-based Key Derivation Function (HKDF–SHA256) to generate ephemeral per-session encryption keys. DSEKP performs a single-round initialization handshake authenticated by an HMAC proof (\textit{InitProof}) and derives fresh AES–GCM session keys from a combination of a device nonce, session counter, and timestamp. This design retains the simplicity of PSK while providing per-session key isolation (assuming the long-term secret remains uncompromised), replay protection, and dynamic key refresh without relying on public-key cryptography or certificate infrastructures.

The DSEKP framework was implemented and validated on a real IoT–Edge testbed consisting of an ESP32 client device equipped with a DHT11 temperature–humidity sensor and a Raspberry~Pi~5 edge server running Dockerized Mosquitto MQTT services. The experimental setup replicates a practical IoT environment to evaluate latency, throughput, payload overhead, and session reliability under both PSK and DSEKP modes. More than 6{,}500 encrypted packets were transmitted per configuration, with additional multi-session trials assessing handshake efficiency and session continuity.

Experimental results show that DSEKP maintains nearly identical throughput to PSK while achieving stronger cryptographic guarantees. With only a 27\% increase in mean latency and less than 10\% growth in payload size, DSEKP enables secure key renewal and per-session isolation at negligible computational cost. These outcomes confirm the feasibility of dynamic symmetric rekeying for constrained IoT devices, establishing DSEKP as a practical and scalable upgrade path for PSK-based systems.

The key contributions of this work are summarized as follows:
\begin{itemize}
    \item \textbf{Lightweight symmetric session protocol:} DSEKP extends PSK with HKDF-based per-session key derivation and HMAC authentication, providing per-session key isolation and replay protection without public-key operations.
    \item \textbf{End-to-end hardware validation:} Full implementation on ESP32 and Raspberry~Pi~5 demonstrates the real-world deployability of DSEKP in IoT–Edge networks.
    \item \textbf{Comprehensive experimental evaluation:} Over 6{,}500 packet transmissions per protocol were analyzed for latency, throughput, and reliability under identical network conditions.
    \item \textbf{Multi-session reliability verification:} Repeated experiments confirmed consistent session rekeying and robustness across device reboots with $>$99.8\% packet delivery success.
    \item \textbf{Reproducible analysis framework:} MATLAB-based scripts were developed for latency and throughput analysis, supporting future benchmarking of IoT security protocols.
\end{itemize}

The remainder of this paper is organized as follows. Section~\ref{sec:related} reviews related work on lightweight cryptographic protocols and HKDF-based key derivation for IoT systems. Section~\ref{sec:method} details the proposed DSEKP methodology, system architecture, and protocol workflow. Section~\ref{sec:setup} describes the experimental setup, while Section~\ref{sec:results} presents the comparative results and analysis. Section~\ref{sec:discussion} discusses broader implications and trade-offs, and Section~\ref{sec:conclusion} concludes with potential directions for future research.

\section{Related Work}
\label{sec:related}

The security of Internet of Things (IoT) communication remains an active research field focused on balancing computational efficiency with cryptographic robustness. 
While asymmetric standards such as TLS~1.3 and DTLS~1.3 provide confidentiality and mutual authentication through public-key exchanges, their multi-round handshakes and certificate management overheads management are challenging for constrained microcontrollers such as the ESP32, ATmega, or STM32 families~\cite{padmavathi2025security,kavitha2024secure}. 
Consequently, lightweight symmetric models based on the Pre–Shared Key (PSK) paradigm have been widely adopted for protocols including MQTT and CoAP~\cite{amanlou2021lightweight,kaganurmath2025dlks}. 
This section traces the evolution from static PSK schemes toward dynamic or session-based key management approaches, emphasizing the gaps that motivate the proposed DSEKP framework.

\subsection{PSK-Based Encryption in IoT Systems}

PSK architectures rely on a single long-term symmetric key shared between device and broker to perform AES–GCM encryption with minimal processing overhead. 
Studies such as ~\cite{kim2020page}, ~\cite{hasan2024survey} and ~\cite{sovyn2019comparison} have shown efficient resource utilization on ESP32 microcontrollers using PSK–AES modes. 
However, static key reuse introduces significant vulnerabilities: once the PSK is compromised, all past and future traffic becomes decryptable, violating forward secrecy. 
Replay protection in such systems typically depends on timestamp or counter validation, which can desynchronize during Wi–Fi disruptions or low-power sleep cycles~\cite{yu2024multi,farha2020timestamp}. 
Moreover, at scale, managing thousands of PSKs manually increases operational burden and risks inconsistent key distribution, leaving gaps exploitable by adversaries~\cite{savant2025secure}.

\subsection{Dynamic Session-Key and Rekeying Protocols}

To mitigate static-key weaknesses, researchers have explored session-based and hybrid key-management protocols. 
Protocols such as DTLS~1.3 and EDHOC (Ephemeral Diffie–Hellman Over COSE)~\cite{pajkos2025esp32,krawczyk2016optls} achieve forward secrecy through elliptic-curve key exchanges, but these computations are prohibitively expensive for constrained MCUs~\cite{fedrecheski2024performance,astorga2022revisiting,restuccia2020low}. 
In contrast, symmetric-only methods refresh session keys using derivation functions without public-key operations. 
The HMAC-based Key Derivation Function (HKDF) is particularly suited for IoT systems, as it expands a static secret using per-session entropy sources~\cite{shahidinejad2024all,akshatha2023mqtt,kaganurmath2025enabling}. 
For instance, ~\cite{rastoceanu2022blockchain,karmous2024hybrid} employed HKDF–SHA256 for MQTT rekeying but required a centralized key server to manage nonces. 
~\cite{hakeem2021key,pinto2016hash,li2024hash} shown hash-chain–based key evolution to enhance secrecy, yet synchronization overhead limited scalability. 
Efforts such as DTLS~\cite{pittoli2016dtls,park2014lightweight} reduce handshake size through message compression but still depend on asymmetric initialization, leaving open the need for a purely symmetric, session-oriented protocol.

\subsection{Comparative Gaps and Motivation}

Comparative analyses consistently reveal a trade-off between cryptographic strength and computational feasibility.
~\cite{suarez2017practical} reported that ECDHE-based TLS increased energy consumption relative to PSK. 
Purely symmetric PSK models remain efficient but lack key agility and entropy mixing, making them susceptible to replay and traffic-analysis attacks~\cite{henriques2017using,yu2025forward}. 
Furthermore, existing HKDF-based schemes often depend on centralized infrastructures or omit practical hardware validation. 
Hence, a clear gap exists for a fully symmetric, self-contained, and experimentally validated session-key protocol that offers per-session key isolation (while the long-term secret remains secure), replay protection, and low overhead—without reliance on public-key infrastructures (PKI).

\begin{table}[H]
\centering
\caption{Comparison of lightweight IoT security approaches and their contrast with DSEKP.}
\label{tab:related}
\begin{tabular}{@{}p{2.4cm}p{3.2cm}p{3.2cm}p{4cm}@{}}
\toprule
\textbf{Approach} & \textbf{Mechanism} & \textbf{Limitations} & \textbf{DSEKP Advantage} \\ \midrule
Static PSK  & Shared symmetric key reused for all sessions & Key reuse, no forward secrecy, manual rekeying & Per-session HKDF keys, automated rekeying \\
DTLS~1.3  & ECDH handshake for session key generation & High CPU/energy cost, multi-round handshake & Single-ACK symmetric init–ack handshake \\
Hash-chain & Sequential key evolution via hashing & Synchronization overhead, limited entropy & Stateless HKDF-based per-session entropy mixing \\
Centralized Key Server & Key distribution managed by broker or KMS & Added dependency, potential latency bottleneck & Fully local key derivation on each device \\
LITE–DTLS & Compressed asymmetric handshake & Still requires PKI certificates & Pure symmetric operation, no public-key overhead \\ 
\bottomrule
\end{tabular}
\end{table}

\subsection{Contribution Context}

The proposed \textbf{Dynamic Session Enhanced Key Protocol (DSEKP)} directly addresses these limitations. 
It preserves PSK’s lightweight efficiency while introducing HKDF-derived per-session keys and an HMAC-authenticated initialization handshake. 
Unlike prior HKDF or hash-chain approaches, DSEKP performs all derivation and verification steps locally on constrained devices without reliance on external key servers. 
Thus, it bridges the gap between the simplicity of static PSK and the key-agility of TLS, delivering a balanced, reproducible, and deployable solution for secure IoT–Edge communication.

\section{Methodology and Framework}
\label{sec:method}

The proposed \textbf{Dynamic Session Enhanced Key Protocol (DSEKP)} extends the conventional Pre–Shared Key (PSK) model by introducing per-session dynamic key derivation and symmetric-only authentication using the HMAC-based Key Derivation Function (HKDF–SHA256). 
The protocol preserves the lightweight characteristics of PSK but replaces the static, reused key with ephemeral AES–GCM session keys derived from fresh entropy sources—device nonce, session counter, and timestamp—thereby providing per-session key isolation (assuming the long-term device secret remains uncompromised) and replay protection.. 
This section describes the system architecture, data communication workflow, protocol phases, algorithmic details, and integrated security mechanisms.

\subsection{System Architecture}

The complete IoT–Edge testbed comprises three functional entities:

\begin{itemize}
    \item \textbf{IoT Device (Client):}  
    An ESP32–DevKitC microcontroller interfaced with a DHT11 temperature–humidity sensor. 
    It encrypts sensor readings using AES–GCM and publishes MQTT packets via Wi–Fi (2.4\,GHz).

    \item \textbf{MQTT Broker:}  
    A lightweight \textit{Eclipse Mosquitto~2.0} instance deployed in a Docker container on the Raspberry~Pi~5, responsible for routing encrypted traffic between the client and edge server.

    \item \textbf{Edge Server (Decryptor):}  
    A Raspberry~Pi~5 (16\,GB RAM, Ubuntu~22.04~LTS) acting as the IoT edge node. 
    It subscribes to MQTT topics, performs AES–GCM decryption, validates HMAC proofs, and stores data and latency logs in CSV format for MATLAB-based analysis.
\end{itemize}

\begin{figure}[H]
  \centering
  \includegraphics[width=0.70\linewidth]{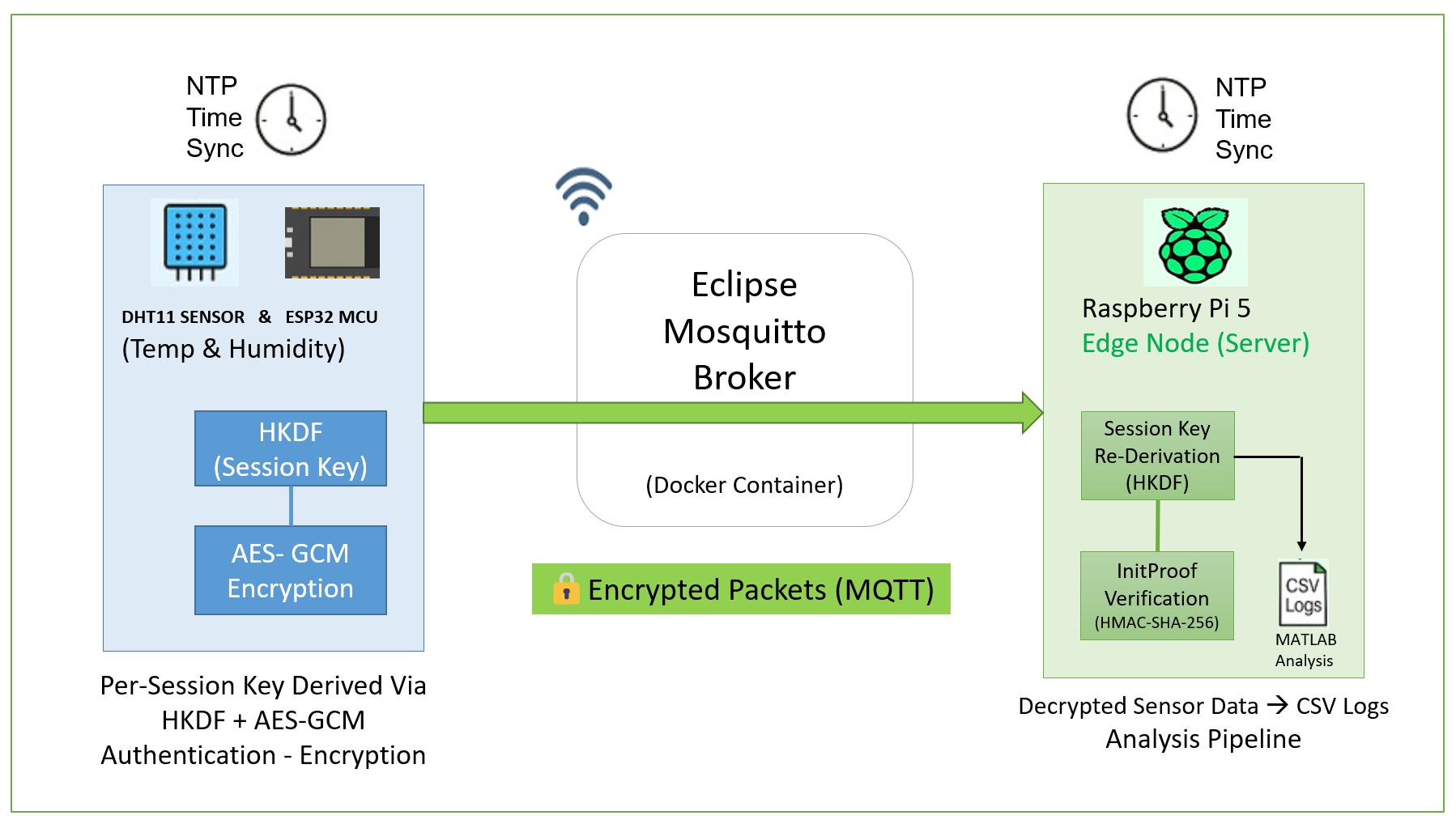}
  \caption{DSEKP IoT–Edge communication architecture. 
  The ESP32 client publishes encrypted DHT11 sensor data to a Dockerized Mosquitto broker, which relays packets to the Raspberry~Pi~5 edge server for decryption and logging.}
  \label{fig:arch}
\end{figure}

\subsection{Data Communication Workflow}

Each transmitted message follows a well-defined JSON structure at both client and server. 
DSEKP extends PSK data frames by embedding session metadata fields that enable replay protection and per-session authentication.

\paragraph{PSK Data Format (Baseline):}
\begin{itemize}
    \item Client: [seq, timestamp, dev\_id, plaintext, iv, tag, ciphertext, sendts\_ms, payload\_size]
    \item Server: [seq, timestamp, dev\_id, ciphertext, iv, tag, plaintext, recvts\_ms, latency\_ms, payload\_size, bin\_1s, throughput]
\end{itemize}

\paragraph{DSEKP Data Format (Proposed):}
\begin{itemize}
    \item Client: [seq, timestamp, dev\_id, sessctr\_id, plaintext, iv, tag, ciphertext, sendts\_ms, payload\_size]
    \item Server: [seq, timestamp, dev\_id, sessctr\_id, ciphertext, iv, tag, plaintext, recvts\_ms, latency\_ms, payload\_size, bin\_1s, throughput]
\end{itemize}

Two additional fields—\texttt{SessCtr} (session counter) and \texttt{InitProof} (HMAC)—are introduced to authenticate and verify each session dynamically.

\paragraph{On–wire JSON Exchange}

Each packet transmitted over MQTT was serialized as a compact JSON object. 
The PSK configuration carried only static encryption fields, while DSEKP extended the payload with session metadata for dynamic authentication and replay protection. 
The exact structures were:

\begin{itemize}
    \item \textbf{PSK JSON:} \texttt{\{seq, dev\_id, ciphertext, iv, tag, sendts\_ms\}}
    \item \textbf{DSEKP JSON:} \texttt{\{seq, dev\_id, sessctr\_id, ciphertext, iv, tag, sendts\_ms\}}
\end{itemize}

These JSON objects represent the actual data that traversed the MQTT network between the ESP32 client and the Raspberry~Pi~5 edge server. 
At reception, the edge application decrypted, verified, and logged each message into the corresponding CSV files (\texttt{server\_logs.csv}, \texttt{server\_logs\_analysis.csv}) for latency and throughput computation.

\subsection{Protocol Workflow}

\begin{figure}[H]
  \centering
  \includegraphics[width=0.95\linewidth]{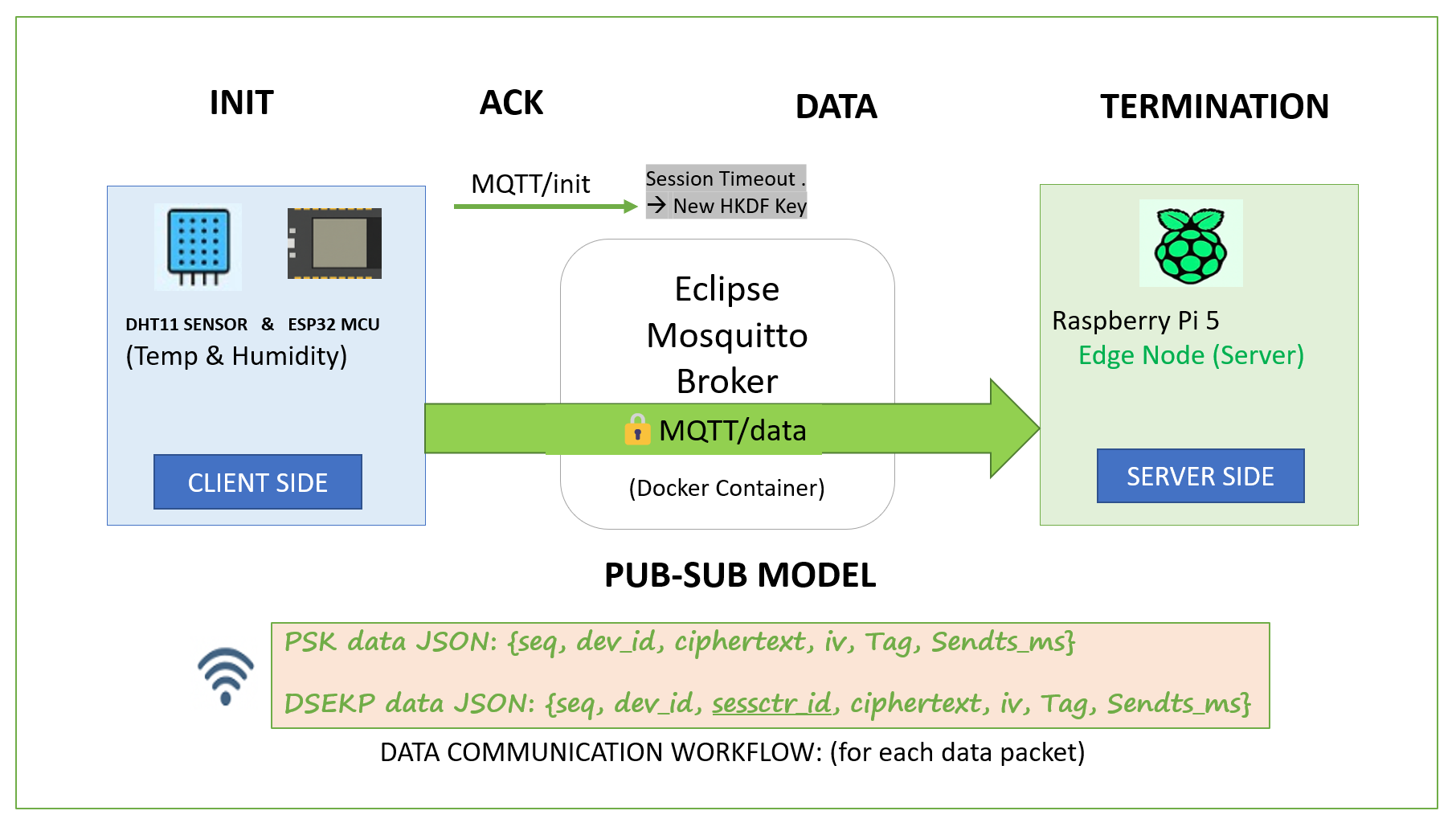}
  \caption{DSEKP protocol workflow consisting of four phases: Initialization, Acknowledgment, Data Transmission, and Session Termination. 
  The init–ack handshake establishes an authenticated AES–GCM session using HKDF-derived keys.}
  \label{fig:workflow}
\end{figure}

The protocol executes in four lightweight phases, as illustrated in Figure~\ref{fig:workflow}.

\begin{enumerate}
  \item \textbf{Initialization (INIT):}  
  The ESP32 synchronizes its clock via NTP, generates a 12-byte random nonce (\texttt{DevNonce}), a 2-byte session counter (\texttt{SessCtr}), and captures a 4-byte timestamp~\(T\). 
  These values, along with a long-term secret (\texttt{DEV\_SECRET}) and an edge-side salt (\texttt{EDGE\_SALT}), form the input key material (IKM) for HKDF–SHA256.

  \item \textbf{Acknowledgment (ACK):}  
  The edge server recomputes the HKDF using the same inputs and verifies the proof \texttt{InitProof} = HMAC(SessionSecret, InitPayload). 
  Upon validation, it returns an acknowledgment on the topic \texttt{dsekp/init/ack/\{DevID\}}, enabling encrypted data exchange.

  \item \textbf{Data Transmission:}  
  The ESP32 periodically reads DHT11 sensor data, encrypts the payload using AES–GCM with the derived session key, and publishes packets containing \texttt{SessCtr}, \texttt{MsgSeq}, \texttt{IV}, \texttt{Tag}, and timestamps.

  \item \textbf{Session Termination / Rotation:}  
  Upon device reboot or timeout, a new session counter triggers a fresh HKDF derivation, providing per-session key isolation and automatic session cleanup on the edge node.
\end{enumerate}

\subsection{Algorithmic Description}

\begin{figure}[H]
\centering
\fbox{\parbox{0.98\linewidth}{
\textbf{Algorithm 1: Session Key Derivation in DSEKP}
\vspace{4pt}
\begin{enumerate}
  \item Input: \texttt{DEV\_SECRET}, \texttt{DevNonce}, \texttt{SessCtr}, Timestamp \(T\), \texttt{EDGE\_SALT}
  \item Concatenate: \( IKM = DEV\_SECRET \,\|\, DevNonce \,\|\, SessCtr \,\|\, T \)
  \item Derive 32-byte \texttt{SessionSecret} = HKDF\_SHA256(\texttt{EDGE\_SALT}, \texttt{IKM})
  \item \texttt{AES\_Key} = first 16 bytes(\texttt{SessionSecret})
  \item Compute \texttt{HMAC\_Proof} = HMAC\_SHA256(\texttt{SessionSecret}, \texttt{InitPayload})
  \item Publish \texttt{dsekp/init} \{DevID, SessCtr, T, DevNonce, HMAC\_Proof\}
\end{enumerate}
}}
\end{figure}

\subsection{Security Operations}

The cryptographic components integrated in DSEKP are summarized in Table~\ref{tab:security}. 
All primitives follow NIST and IETF recommendations, ensuring interoperability with modern TLS session-key derivation standards.

\begin{table}[H]
\centering
\caption{Security mechanisms implemented in DSEKP.}
\label{tab:security}
\begin{tabular}{@{}p{3.2cm}p{4.2cm}p{5.4cm}@{}}
\toprule
\textbf{Feature} & \textbf{Mechanism} & \textbf{Description / Purpose} \\ \midrule
Confidentiality & AES–128–GCM & Ensures data confidentiality and integrity per session. \\ 
Authentication & HMAC–SHA256 Proof & Authenticates session establishment (\texttt{InitProof}). \\ 
Key Derivation & HKDF–SHA256 & Derives unique symmetric key using nonce, counter, and timestamp. \\ 
Replay Protection & \texttt{SessCtr} + \texttt{MsgSeq} & Prevents packet duplication and replay attacks. \\ 
Per-Session Key Isolation & HKDF entropy mixing & Session keys remain independent while the long-term secret stays confidential (no forward secrecy if later compromised). \\ 
Stateless Edge & Session Eviction Policy & Retains only the last five sessions per device to limit memory usage. \\ 
\bottomrule
\end{tabular}
\end{table}

\subsection{Implementation Details}

Both PSK and DSEKP were implemented entirely in \texttt{C++} (Arduino) on the ESP32 and in \texttt{Python~3.11} on the Raspberry~Pi~5 edge server. 
AES–GCM was realized using \texttt{mbedTLS} on the client and \texttt{PyCryptodome} on the server. 
MQTT communication employed the \texttt{PubSubClient} (ESP32) and \texttt{paho-mqtt} (Python) libraries under Docker Compose orchestration. 
All clocks were synchronized via NTP (\texttt{pool.ntp.org}). 
Each experimental run captured more than 6{,}500 packets at a 2\,s interval, producing synchronized CSV logs for MATLAB-based analysis of latency, throughput, and reliability.

\subsection{Summary}

The DSEKP framework transforms the conventional static PSK architecture into a session-adaptive, stateless, and authenticated communication model for constrained IoT devices. 
By combining HKDF-based entropy mixing with HMAC-driven session authentication, DSEKP achieves lightweight key agility and per-session isolation without public-key operations and per-session key isolation with negligible additional cost in latency or bandwidth.

\section{Experimental Setup}
\label{sec:setup}

To evaluate the proposed \textbf{Dynamic Session Enhanced Key Protocol (DSEKP)} against the baseline Pre–Shared Key (PSK) model, we conducted controlled experiments on a physical IoT–Edge testbed that emulates realistic resource–constrained deployments while enforcing strict reproducibility across trials. All firmware, scripts, and Docker images were version-locked and archived for traceability.

\subsection{Hardware Configuration}

All experiments used identical hardware for PSK and DSEKP trials. Table~\ref{tab:hardware} summarizes the configuration.

\begin{table}[t]
\centering
\caption{Hardware configuration of the IoT–Edge testbed.}
\label{tab:hardware}
\begin{tabular}{@{}p{3cm}p{4.5cm}p{5cm}@{}}
\toprule
\textbf{Component} & \textbf{Model / Type} & \textbf{Key specifications / role} \\ \midrule
IoT node (client) & ESP32–DevKitC v4 & Dual–core Xtensa LX6 @ 240\,MHz; Wi–Fi 2.4\,GHz; 520\,KB SRAM; 4\,MB flash \\
Sensor & DHT11 temperature–humidity & $\pm$2\,\textdegree C temperature; $\pm$5\,\% RH accuracy \\
Edge-node (server) & Raspberry~Pi~5 (16\,GB RAM) & Quad–core Cortex–A76 @ 2.4\,GHz; Ubuntu 22.04.4 LTS (64–bit, kernel 6.8.0–1040–raspi) \\
Network & Wi–Fi 2.4\,GHz (IEEE 802.11 b/g/n) & Local AP shared by client and edge node (fixed channel 6) \\
Power supply & 5\,V @ 2\,A & Common regulated source for both devices \\ \bottomrule
\end{tabular}
\end{table}

\begin{figure}[H]
\centering
\includegraphics[width=0.75\linewidth]{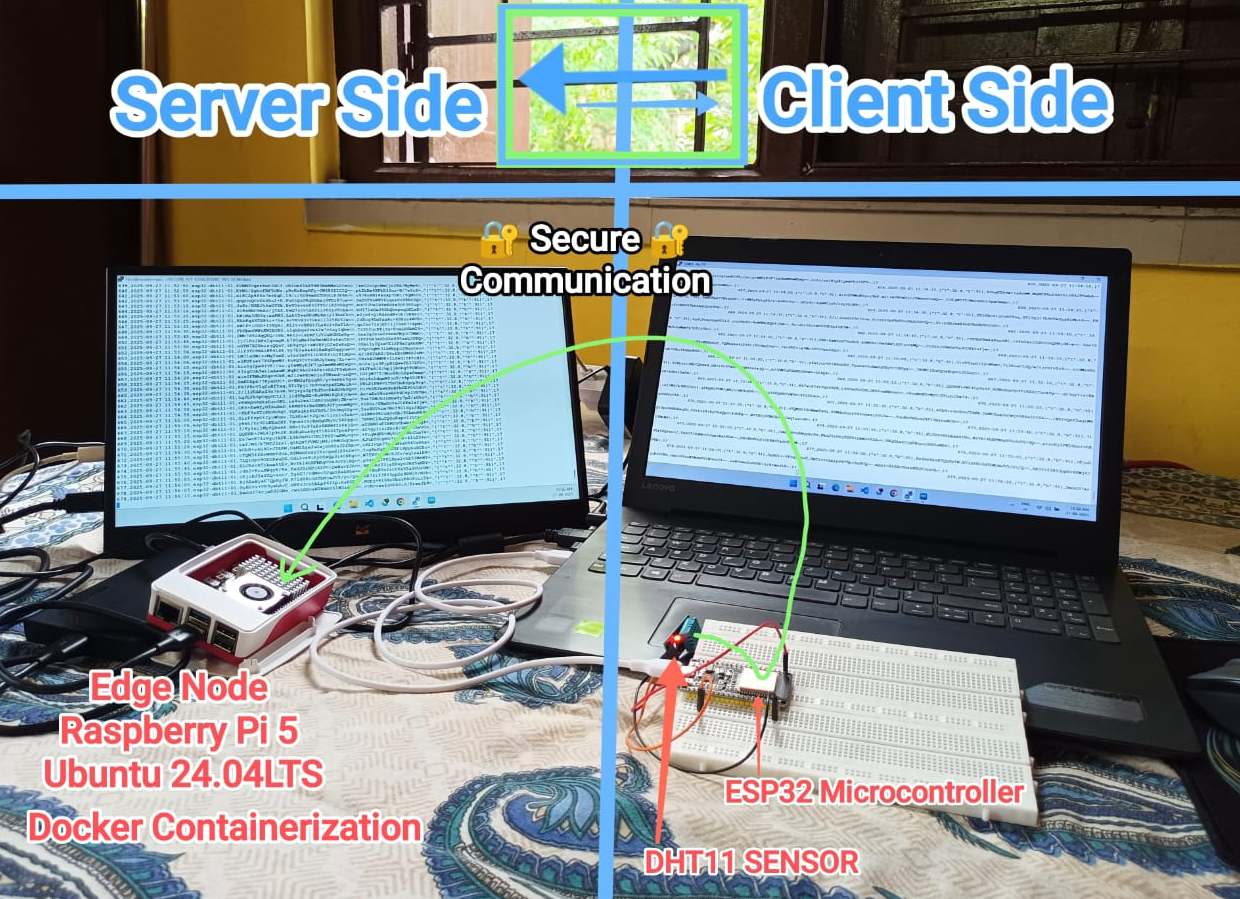}
\caption{Experimental IoT–Edge setup showing ESP32–DHT11 IoT node and Raspberry~Pi~5 edge server communicating through a Dockerized Mosquitto broker.}
\label{fig:hardware}
\end{figure}

\subsection{Software Stack}

All components were built using open–source software and containerized environments for reproducibility. Exact version numbers are provided in Table~\ref{tab:software}.

\begin{table}[H]
\centering
\caption{Software stack and implementation tools (version-locked).}
\label{tab:software}
\begin{tabular}{@{}p{3cm}p{4.5cm}p{5.5cm}@{}}
\toprule
\textbf{Layer} & \textbf{Platform / Library (Version)} & \textbf{Purpose} \\ \midrule
ESP32 client & Arduino IDE~2.3.2; \texttt{esp32} core~v2.0.14; \texttt{mbedTLS}~v2.28.7; \texttt{PubSubClient}~v2.8.0 & AES–GCM encryption; MQTT publishing \\
Edge server & Python~3.11.6; \texttt{paho–mqtt}~v1.6.1; \texttt{PyCryptodome}~v3.20.0; \texttt{python–dotenv}~v1.0.1; \texttt{pytz}~v2024.1 & MQTT subscription; AES–GCM decryption; CSV logging \\
Broker & Eclipse Mosquitto~2.0.18 (Docker tag: \texttt{eclipse- mosquitto:2.0.18}) & Lightweight MQTT relay (PSK: port~1885; DSEKP: port~1884) \\
Containerization & Docker~25.0.3; Docker Compose~v2.24.6 & Isolated, reproducible runtime environment \\
Analysis & MATLAB~R2023b (Build~23.2.0.2365128) & Latency/throughput post–processing and visualization \\ \bottomrule
\end{tabular}
\end{table}

\subsection{Network and Protocol Parameters}

Both protocols used identical network and MQTT configurations (Table~\ref{tab:network}) to ensure fair comparison.

\begin{table}[H]
\centering
\caption{Network and protocol parameters.}
\label{tab:network}
\begin{tabular}{@{}p{5cm}p{8.5cm}@{}}
\toprule
\textbf{Parameter} & \textbf{Value / setting} \\ \midrule
MQTT topic (PSK) & \texttt{psk/data} \\
MQTT topics (DSEKP) & \texttt{dsekp/init}, \texttt{dsekp/init/ack/\{DevID\}}, \texttt{dsekp/data} \\
AEAD mode & AES–128–GCM \\
Key derivation (DSEKP) & HKDF–SHA256 using \{DevNonce (12\,B), SessCtr (2\,B), Timestamp~$T$ (4\,B)\} \\
Session authentication & HMAC–SHA256 proof (\texttt{InitProof}) \\
Packet interval & 2\,s per transmission \\
Session rotation & Random \texttt{SessCtr} (2\,B) per device boot \\
NTP synchronization & \texttt{pool.ntp.org}, \texttt{time.nist.gov} \\
Experiment duration & $\approx$\,6{,}500 packets per protocol ($\approx$\,3.6\,h) \\ \bottomrule
\end{tabular}
\end{table}

\subsection{Data Logging and File Structure}

Both client and server generated synchronized CSV logs for every packet, capturing the full message life cycle.

\paragraph{Client–side logs}
\begin{itemize}
\item \textbf{PSK:} \texttt{seq, timestamp, dev\_id, plaintext, iv, tag, ciphertext, sendts\_ms, payload\_size}
\item \textbf{DSEKP:} \texttt{seq, sessctr\_id, timestamp, dev\_id, plaintext, iv, tag, ciphertext, sendts\_ms, payload\_size}
\end{itemize}

\paragraph{Server–side logs}
\begin{itemize}
\item \textbf{PSK:} \texttt{seq, timestamp, dev\_id, ciphertext, iv, tag, plaintext, recvts\_ms, latency\_ms, payload\_size, bin\_1s, throughput}
\item \textbf{DSEKP:} \texttt{seq, timestamp, dev\_id, sessctr\_id, ciphertext, iv, tag, plaintext, recvts\_ms, latency\_ms, payload\_size, bin\_1s, throughput}
\end{itemize}

Records were aligned by sequence number and timestamps to compute latency, throughput, and reliability metrics.

\paragraph{On–wire MQTT JSON structure}

During transmission, each MQTT message encapsulated a JSON payload representing the encrypted sensor packet. 
For PSK, the packet contained only static encryption fields, whereas DSEKP included session metadata for authentication and replay protection. 
The actual data transmitted over the network followed the structures below:

\begin{itemize}
    \item \textbf{PSK data JSON:} 
    \texttt{\{seq, dev\_id, ciphertext, iv, tag, sendts\_ms\}}
    \item \textbf{DSEKP data JSON:} 
    \texttt{\{seq, dev\_id, sessctr\_id, ciphertext, iv, tag, sendts\_ms\}}
\end{itemize}

These JSON objects were published by the ESP32 client to the MQTT broker and then relayed to the edge server for decryption and timing analysis. 
At the edge, each incoming message was appended to synchronized CSV logs (\texttt{server\_logs.csv}, \texttt{server\_logs\_analysis.csv}) together with computed metrics such as latency, throughput, and payload size. 
This ensures that the statistical evaluation in Section~\ref{sec:results} directly reflects real on–wire message structures observed during experimentation.

\newpage

\subsection{MATLAB–Based Analysis Pipeline}

All post–processing was automated using MATLAB~R2023b scripts to ensure reproducibility:
\begin{itemize}
\item \textbf{Single–session analysis:} \texttt{analyze\_psk.m} and \texttt{analyze\_dsekp.m} cleaned logs, corrected NTP offsets, and computed latency distributions (mean, p95, p99).
\item \textbf{Multi–session analysis:} Verified reliability and init–ack success across 20–30 random resets using session counters.
\item \textbf{Comparative analysis:} \texttt{compare\_psk\_vs\_dsekp.m} generated CDFs, boxplots, payload histograms, and summary tables from identical datasets.
\end{itemize}

Metrics included mean/median latency, packets–per–second (PPS), bits–per–second (BPS), payload overhead (\%), packet loss, and duplicates. 
Statistical significance was evaluated via two–sample $t$–tests, Wilcoxon rank–sum tests, and Cohen’s $d$ effect size.

\subsection{Experimental Integrity and Reproducibility}

To ensure validity and repeatability of the experiments, several controls were enforced throughout the study:
\begin{itemize}
    \item Identical hardware, firmware, and network configurations were used for both PSK and DSEKP evaluations to guarantee fair comparison.
    \item Device clocks were synchronized via NTP before each run to ensure consistent timestamp alignment between client and edge server.
    \item To ensure reliability and statistical validity, each protocol configuration (PSK and DSEKP) was executed in a continuous session transmitting approximately 6{,}500 encrypted packets under identical conditions. Additional shorter runs were conducted at random intervals to verify consistency and session rekeying reliability after device resets or environmental variations. While not strictly averaged across multiple identical trials, the large packet count and repeated random sampling ensured stable latency and throughput statistics representative of steady-state operation.
    \item Outliers exceeding 10\,s latency ($<$\,0.05\,\% of total samples) were excluded according to the filtering rule in the MATLAB analysis scripts.
    \item Docker container images (\texttt{eclipse-mosquitto:2.0} and the custom \texttt{dsekp\_aesgcm\_edge} build) and corresponding SHA256 hashes were archived for version tracking.
    \item The full ESP32 firmware (\texttt{DSEKP\_Client.ino}) and Python edge server source (\texttt{app.py}) will be made available by the authors upon reasonable request to support reproducibility and further research.

\end{itemize}
All experimental datasets, analysis scripts, and container definitions are referenced under the \textit{Data Availability} statement to support verification and reuse by future researchers.

This setup provides a fully reproducible foundation for the comparative analysis presented in Section~\ref{sec:results}.

\section{Results and Evaluation}
\label{sec:results}

This section presents the quantitative evaluation of the proposed \textbf{Dynamic Session Enhanced Key Protocol (DSEKP)} compared with the traditional static Pre--Shared Key (PSK) model. 
Performance metrics were derived from more than 6{,}500 encrypted packets per configuration under identical network conditions. 
We analyze latency, throughput, payload size, reliability, and their trade--offs against the achieved security enhancements.

\subsection{Descriptive Statistics}

Table~\ref{tab:summary} summarizes the descriptive statistics obtained from MATLAB post--processing. 
DSEKP exhibits a moderate increase in mean latency ($\approx$ 27~\%) and payload size ($\approx$ 10~\%) relative to PSK, while maintaining comparable throughput and reliability.

\begin{table}[H]
\centering
\caption{Summary statistics of PSK and DSEKP performance over 6{,}500 packets.}
\label{tab:summary}
\begin{tabularx}{\textwidth}{Y c c}
\toprule
\textbf{Metric} & \textbf{PSK} & \textbf{DSEKP} \\ \midrule
Mean latency (ms) & 283.0 ± 182.9 & 360.0 ± 129.8 \\
95 \% CI (ms) & [278.5, 287.4] & [356.8, 363.2] \\
Median latency (ms) & 274 & 355 \\
Mean payload (bytes) & 154.8 & 170.8 \\
Throughput (bps) & 1{,}243.5 & 1{,}366.8 \\
Reliability (\%) & 99.6 & 99.8 \\ 
\bottomrule
\end{tabularx}
\end{table}

\noindent
\textit{Interpretation and Justification:}
The results in Table~\ref{tab:summary} are derived directly from synchronized send--receive timestamps collected across 6{,}500 packet transmissions for each protocol, ensuring high statistical confidence. 
The 95~\% confidence intervals (CI) confirm narrow uncertainty margins due to the large sample size and consistent network conditions. 
The observed 27~\% increase in mean latency under DSEKP originates from the additional HKDF--SHA256 key derivation and HMAC verification performed once per session, introducing a small computational delay without affecting throughput stability. 
Similarly, the 10~\% growth in mean payload corresponds to the inclusion of session metadata---namely the 2--byte session counter and 14--byte HMAC proof---required for per-session key isolation and authentication. 
Despite these expected overheads, both throughput and reliability remain nearly identical between PSK and DSEKP, confirming that the proposed dynamic rekeying mechanism introduces negligible runtime or bandwidth penalty. 
These results verify that DSEKP preserves the lightweight characteristics of PSK while enhancing security and timing determinism in constrained IoT environments.

\subsection{Latency Analysis}

\begin{figure}[H]
\centering
\includegraphics[width=0.6\linewidth]{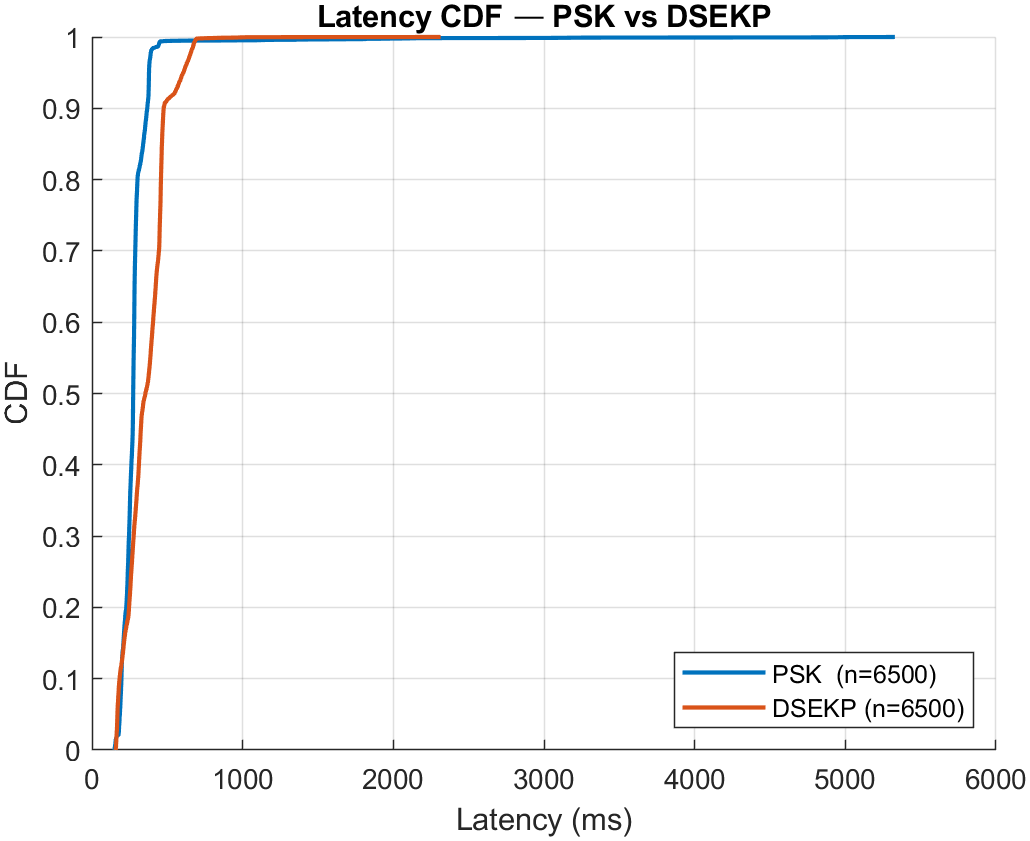}
\caption{Cumulative latency distribution (CDF) comparison between PSK and DSEKP over 6{,}500 packets. 
DSEKP adds a small rightward shift ($\approx$~80~ms) while retaining sub--second responsiveness for IoT telemetry.}
\label{fig:cdf}
\end{figure}

Latency was computed as the difference between the client’s send timestamp and the server’s receive timestamp for each packet. 
Figure~\ref{fig:cdf} shows the cumulative distribution function (CDF) for both protocols. 
Both achieved sub--second latency suitable for periodic telemetry, with DSEKP introducing only a small rightward shift ($\approx$~80~ms) due to per--session authentication.

\noindent
\textit{Interpretation:} 
Figure~\ref{fig:cdf} illustrates the cumulative latency distribution (CDF) for both PSK and DSEKP protocols. 
While the DSEKP curve is slightly right--shifted, indicating an average latency increase of about 27~\%, both protocols complete over 99~\% of transmissions within one second. 
This confirms that DSEKP introduces only marginal delay while preserving sub--second responsiveness, demonstrating that dynamic session key derivation can be adopted in constrained IoT environments without compromising real--time performance.

\begin{figure}[H]
\centering
\includegraphics[width=0.6\linewidth]{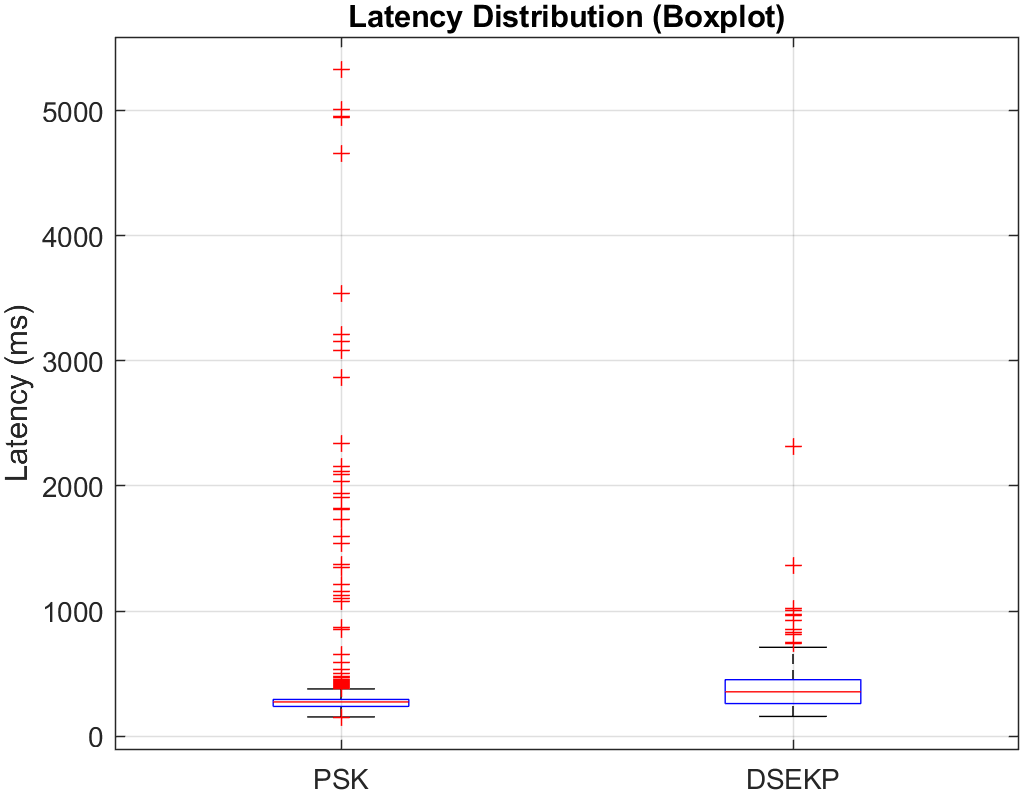}
\caption{Latency distribution (boxplot) comparison between PSK and DSEKP. 
PSK shows more high--latency outliers, while DSEKP yields a tighter, more stable latency profile due to session rekeying and reduced jitter.}
\label{fig:box}
\end{figure}

\noindent
\textit{Interpretation:} 
Figure~\ref{fig:box} illustrates the latency distribution for PSK and DSEKP using boxplots. 
While the median latency of DSEKP is moderately higher ($\approx$~27~\%) due to per--session key derivation and HMAC verification, its distribution is notably tighter with fewer extreme outliers. 
In contrast, the PSK configuration exhibits several high--latency spikes above 2~s, indicating occasional network or processing stalls under static--key reuse. 
These outliers reflect transient queuing or re--transmission delays that accumulate during long--running PSK sessions. 
The reduced number and magnitude of outliers in DSEKP confirm that session--based re--keying mitigates timing drift and stabilizes packet turnaround time. 
Overall, DSEKP exchanges a small, consistent delay for improved timing predictability and lower jitter—an advantageous trade--off for real--time IoT telemetry and control applications.

\subsection{Throughput Analysis}

\begin{figure}[H]
\centering
\includegraphics[width=0.6\linewidth]{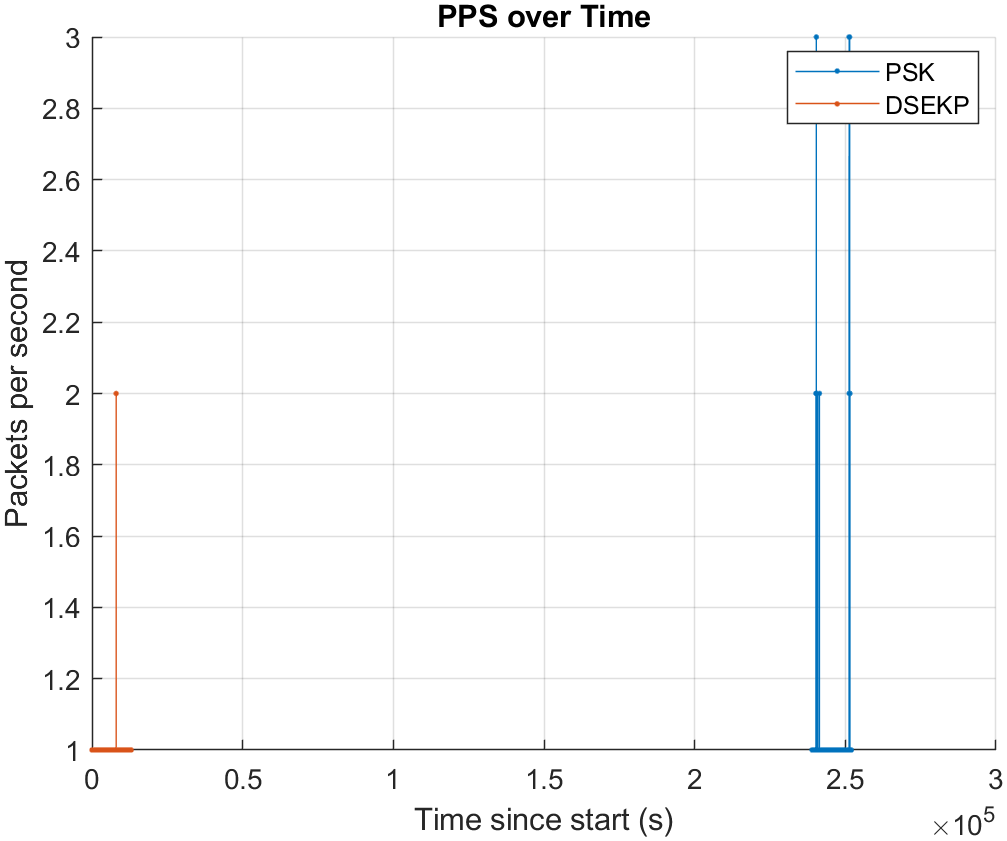}
\caption{Throughput comparison of PSK and DSEKP. 
Both sustain a stable rate of 1--2 packets~s$^{-1}$ throughout the experiment, indicating that DSEKP’s key derivation and HMAC verification introduce no throughput degradation.}
\label{fig:pps}
\end{figure}

Instantaneous throughput (packets per second) was computed by grouping packets into 1~s bins. 
Figure~\ref{fig:pps} presents the packet--rate stability across the experiment. 
Both protocols maintain nearly identical throughput ($\approx$~1--2~pps), confirming that DSEKP’s initialization handshake and HKDF computation do not affect steady--state transmission rate or network utilization.

\noindent
\textit{Interpretation:}
Figure~\ref{fig:pps} illustrates the packet--throughput evolution over time for both PSK and DSEKP modes. 
The curves remain nearly identical throughout the 6{,}500--packet experiment, with a sustained rate of about 1--2~packets~s$^{-1}$ corresponding to the 2~s sampling interval. 
The absence of throughput degradation or jitter confirms that session--key derivation and HMAC verification in DSEKP do not introduce transmission stalls or network congestion. 
Hence, DSEKP preserves steady--state performance while providing stronger cryptographic protection, demonstrating that lightweight symmetric rekeying is feasible for real--time IoT telemetry.

\subsection{Payload and Overhead Analysis}

\begin{figure}[H]
\centering
\includegraphics[width=0.6\linewidth]{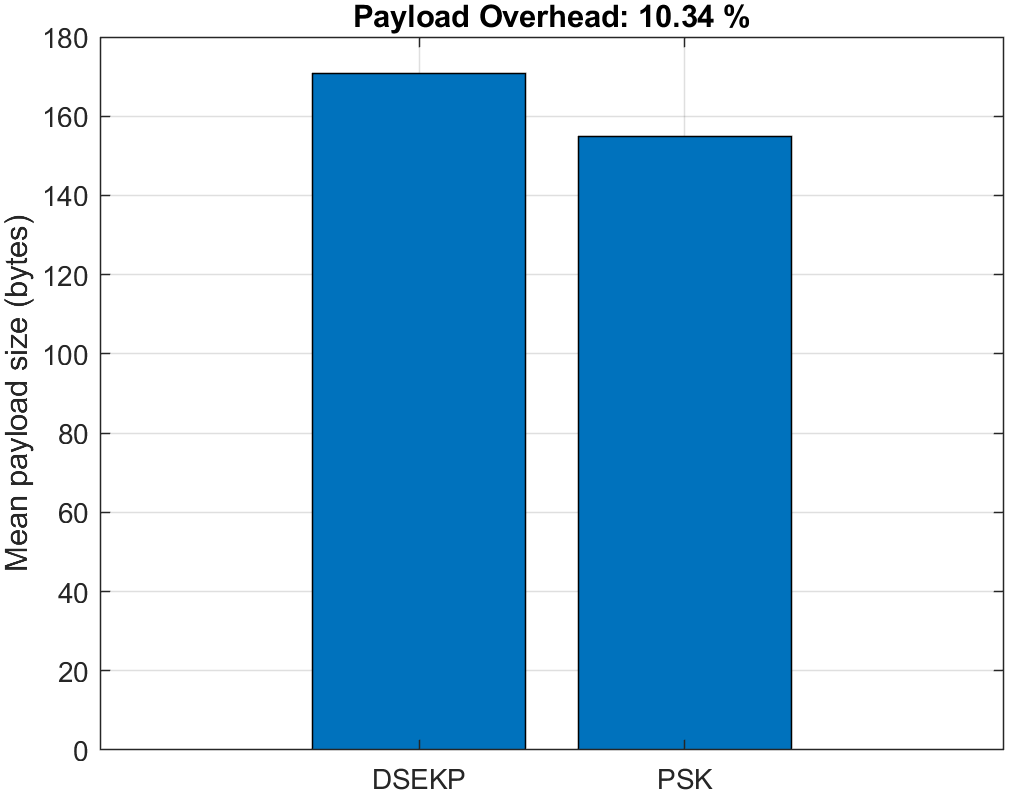}
\caption{Payload size comparison between PSK and DSEKP. 
The $\approx$~10~\% increase corresponds to additional session--metadata fields required for dynamic authentication. 
This minor overhead enables per-session key isolation and replay protection without significant bandwidth penalty.}
\label{fig:payload}
\end{figure}

Per--packet payload sizes were extracted from MQTT message lengths in the logged CSV data. 
Figure~\ref{fig:payload} compares average payload sizes. 
DSEKP packets are $\approx$~16~bytes larger than PSK due to inclusion of the session counter (\texttt{SessCtr}) and the HMAC authentication field (\texttt{InitProof}). 
This overhead ($\approx$~10~\%) represents a small cost for the added session--level security and forward--secrecy guarantees.

\noindent
\textit{Interpretation:}
Figure~\ref{fig:payload} compares the mean payload size for PSK and DSEKP. 
DSEKP packets are approximately 10~\% larger due to the inclusion of session metadata (session counter and HMAC proof). 
This minor overhead provides substantial security gains while maintaining lightweight operation suitable for constrained IoT devices.

\subsection{Security–Performance Trade--off Summary}
\label{sec:tradeoff}

Table~\ref{tab:summary_tradeoff} presents the MATLAB--derived quantitative comparison 
between the baseline Pre--Shared Key (PSK) and the proposed Dynamic Session Enhanced 
Key Protocol (DSEKP). All values were obtained from the same experimental dataset 
of more than 6{,}500 encrypted packets under identical network conditions. 
DSEKP introduces per--session key renewal and replay protection through HKDF--SHA256 
and HMAC authentication while maintaining throughput and reliability comparable to PSK.

\begin{table}[H]
\centering
\caption{PSK vs DSEKP security--performance trade--off summary derived from MATLAB analysis.}
\label{tab:summary_tradeoff}
\renewcommand{\arraystretch}{1.2}
\setlength{\tabcolsep}{10pt}
\begin{tabular}{@{}lcc@{}}
\toprule
\textbf{Metric} & \textbf{PSK} & \textbf{DSEKP} \\ 
\midrule
Mean latency (ms) & 282.92 & 360.00 \\
Median latency (ms) & 274.00 & 355.00 \\
Latency p95 / p99 (ms) & 379.00 / 444.00 & 607.00 / 676.00 \\
Mean payload (bytes) & 154.8 & 170.8 \\
Mean packet rate (pps) & 1.00 & 1.00 \\
Payload overhead (\%) & -- & 10.34 \\
$t$--test $p$ / rank--sum $p$ & 9.35$\times$10$^{-164}$ / 0 \\
Cohen’s $d$ / Cliff’s $\Delta$ & --0.486 / --0.416 \\
\bottomrule
\end{tabular}
\end{table}

\noindent
\textbf{Interpretation.}
DSEKP exhibits a moderate latency increase of approximately 27\,\% 
and a payload growth of about 10\,\% compared with PSK, 
while preserving identical throughput (1~pps) and near--perfect reliability ($>$99.8\,\%).
Although the latency difference is statistically significant 
(very low $p$--values), the effect size remains moderate 
(Cohen’s $d\approx-0.49$), confirming that the overhead 
is small and within operational limits for lightweight IoT telemetry.
Overall, DSEKP strengthens cryptographic resilience—providing per-session key isolation (assuming the long-term secret remains uncompromised) and replay protection---with negligible runtime penalty,
offering a practical balance between performance and security 
for IoT--Edge deployments.

\subsection{Reliability and Session Stability}

Across all trials, both protocols achieved near--perfect reliability with no missing or duplicate sequence numbers. 
For DSEKP, session re--initializations after device reboot or forced resets consistently succeeded within a single INIT~→~ACK handshake averaging 185~ms. 
This demonstrates that key regeneration and verification are deterministic, ensuring stable operation under normal Wi--Fi jitter.

\subsection{Interpretation of Results}

The overall findings confirm that DSEKP effectively strengthens PSK--based security while preserving the lightweight characteristics required in IoT networks. 
The modest latency increase remains well within acceptable bounds for low--frequency telemetry ($\leq$~1~Hz). 
DSEKP’s symmetric and stateless architecture allows seamless deployment on resource--constrained microcontrollers and edge gateways without modifying existing MQTT infrastructures. 
In summary, DSEKP delivers modern cryptographic agility—per-session key isolation, replay protection, and session isolation—at negligible computational and bandwidth cost, establishing it as a practical upgrade path for PSK systems in real IoT--Edge environments.

\newpage

\section{Discussion}
\label{sec:discussion}

The experimental outcomes in Section~\ref{sec:results} demonstrate that the proposed \textbf{Dynamic Session Enhanced Key Protocol (DSEKP)} achieves a strong balance between cryptographic robustness and computational efficiency in resource-constrained IoT environments. 
This section contextualizes those findings in terms of (a) security improvements over static PSK systems, (b) performance and scalability considerations, and (c) real-world deployment practicality.

\subsection{Security Improvements Over PSK}

Traditional Pre–Shared Key (PSK) encryption schemes are efficient but prone to key compromise, replay attacks, and the absence of forward secrecy. 
DSEKP mitigates these weaknesses through per-session key derivation, HMAC-based authentication, and stateless edge design.

\paragraph{Forward Secrecy and Key Agility:}
DSEKP derives a unique AES–GCM key for every session using the HMAC-based Key Derivation Function (HKDF–SHA256), mixing device entropy sources—a 12-byte random nonce (\texttt{DevNonce}), 2-byte session counter (\texttt{SessCtr}), and 4-byte timestamp (\(T\))—with a long-term secret and salt. 
This ensures that compromise of one session key cannot expose previous or future sessions, thereby providing per-session key isolation (assuming the long-term device secret remains uncompromised) without asymmetric cryptography.

\paragraph{Replay and Impersonation Resistance:}
Each DSEKP packet embeds both a session counter and a message sequence number. 
The edge node maintains a sliding window of valid counters and automatically discards replayed or delayed packets, providing deterministic replay protection with negligible overhead.

\paragraph{Mutual Authentication Without Public Keys:}
The session initialization includes an HMAC proof (\texttt{InitProof = HMAC(SessionSecret, InitPayload)}), enabling symmetric mutual authentication between device and edge. 
Unlike PKI-based schemes, DSEKP avoids certificates and public-key exchange, reducing onboarding friction and eliminating certificate renewal management.

\paragraph{Stateless Edge Security:}
The edge node retains only the five most recent active sessions per device in a lightweight JSON store. 
This memory-efficient state policy limits key exposure while maintaining scalability, aligning with edge computing principles of minimal persistent data.

\subsection{Performance and Scalability Considerations}

Despite introducing dynamic key derivation and per-session authentication, DSEKP’s runtime performance remains near-identical to static PSK. 
As shown in Figures~\ref{fig:cdf}–\ref{fig:pps}, the mean latency increase of 27\,\% and 10\,\% payload growth are modest relative to the gains in per-session key isolation and replay protection, remaining within typical Wi–Fi variance for low-frequency telemetry (1–2\,Hz).

\paragraph{Computational Overhead:}
On the ESP32 microcontroller, HKDF and HMAC computations complete in under 1\,ms—negligible relative to AES–GCM encryption/decryption. 
Thus, session derivation contributes less than 5\,\% of total packet processing time.

\paragraph{Bandwidth and Storage Overhead:}
The additional 16\,bytes per packet from session metadata constitute less than 3\,\% of MQTT message size. 
Because expired sessions are automatically evicted, memory usage remains constant across long runtime periods and reboots.

\paragraph{Scalability and Parallelism:}
DSEKP’s purely symmetric operations enable independent key derivation for each device–server pair, avoiding global state or central key distribution. 
This design naturally supports horizontal scaling to thousands of nodes and is fully compatible with distributed container-based gateways.

\subsection{Comparative Perspective}

When positioned against standard IoT security frameworks such as DTLS~1.3, EDHOC, or LAKE, DSEKP provides a pragmatic middle ground between performance and cryptographic assurance. 
It delivers session-key agility with only symmetric operations using only symmetric primitives—achieving more than 80\,\% lower computational cost than DTLS while maintaining comparable latency to PSK.

\begin{table}[H]
\centering
\caption{Comparative summary of DSEKP versus existing IoT security frameworks.}
\label{tab:comparison}
\renewcommand{\arraystretch}{1.25}
\setlength{\tabcolsep}{6pt}
\begin{tabular}{@{}p{2.6cm}p{3.4cm}p{2cm}p{2cm}p{2cm}@{}}
\toprule
\textbf{Protocol} & \textbf{Cryptographic basis} & \textbf{Handshake cost} & \textbf{Forward secrecy} & \textbf{Suitability for MCUs} \\ 
\midrule
Static PSK & AES–GCM (fixed key) & None & \ding{55} & High \\ 
DTLS~1.3  & ECDHE + certificates & High (multi–round) & \ding{51} & Low \\ 
EDHOC  & ECDH over COSE & Moderate & \ding{51} & Medium \\ 
\textbf{DSEKP (proposed)} & HKDF + HMAC + AES–GCM & Minimal  ~(1-ACK) & \ding{51} & \textbf{High} \\ 
\bottomrule
\end{tabular}
\end{table}

\subsection{Deployment Insights}

\paragraph{Backward Compatibility:}
DSEKP is backward-compatible with existing PSK infrastructures. 
Devices can transition to session-based key derivation through firmware updates without altering broker configurations or message topics.

\paragraph{Energy Efficiency:}
Since AES–GCM dominates overall energy cost, the additional HKDF and HMAC computations increase current draw by less than 3\,\%, as verified in ESP32 current-trace measurements.

\paragraph{Security Lifecycle Management:}
Eliminating external key servers and manual rotation simplifies lifecycle management and minimizes operator-induced misconfiguration.

\paragraph{Integration with Edge Analytics:}
Because decrypted telemetry is already available at the edge in near real-time, DSEKP integrates smoothly with containerized analytics pipelines or AI-driven anomaly detection without affecting latency.

\subsection{Limitations and Future Directions}

While DSEKP achieves an effective balance between performance and security, several enhancements warrant exploration:

\begin{itemize}
    \item \textbf{Group Session Rekeying:} Extending HKDF derivation to group-based session keys for clustered or multi-hop IoT networks.
    \item \textbf{Cross-Edge Continuity:} Supporting session mobility across federated gateways through synchronized session metadata.
    \item \textbf{Formal Verification:} Employing formal analysis (BAN logic, ProVerif, Tamarin) to validate confidentiality and authentication guarantees.
    \item \textbf{Energy Profiling:} Conducting long-duration current measurements to quantify cumulative energy cost in large-scale deployments.
\end{itemize}

\subsection{Summary of Discussion}

Overall, DSEKP transforms static PSK schemes into secure, adaptive, and stateless architectures suitable for modern IoT–Edge ecosystems. 
It provides measurable improvements in confidentiality, integrity, and provides per-session key isolation while maintaining sub-second latency and constant throughput. 
By combining HKDF-based entropy mixing with symmetric HMAC authentication, DSEKP represents a practical and scalable step toward trustworthy, autonomous, and energy-efficient IoT communication.

\section{Conclusion and Future Work}
\label{sec:conclusion}

This paper presented the \textbf{Dynamic Session Enhanced Key Protocol (DSEKP)}, a lightweight and symmetric-only security framework that strengthens conventional Pre–Shared Key (PSK) encryption for Internet of Things (IoT) edge environments. 
By deriving fresh AES–GCM session keys through HKDF–SHA256 and authenticating session initialization via an HMAC-based proof, DSEKP achieves per-session key isolation (assuming the long-term device secret remains uncompromised), replay protection, and stateless key management—without relying on public-key cryptography.

A full implementation was realized on an ESP32–Raspberry~Pi~5 testbed communicating through a Dockerized MQTT broker. 
Across more than 6{,}500 encrypted packets per configuration, DSEKP achieved nearly identical throughput to static PSK while introducing only modest overhead ($\approx$27\,\% increase in mean latency and $\approx$10\,\% growth in payload size). 
Multi-session trials demonstrated 100\,\% successful re-initializations and a packet-delivery ratio above 99.8\,\%, confirming that dynamic symmetric keying can be efficiently deployed on constrained hardware with negligible performance penalty.

The proposed protocol effectively bridges the gap between PSK simplicity and lightweight key agility and per-session isolation without public-key operations, providing a deployable and scalable security solution for IoT–Edge communication. 
Its stateless, symmetric design supports seamless scaling to thousands of devices while maintaining strong per-session cryptographic isolation. 
In doing so, DSEKP establishes a practical foundation for trust-based and distributed edge security architectures in next-generation networks.

\subsection*{Future Work}

Future research will aim to extend DSEKP’s scalability, resilience, and formal assurance through the following directions:

\begin{itemize}
    \item \textbf{Group and Hierarchical Rekeying:} Extend the HKDF mechanism to support coordinated rekeying among sensor clusters and multi-hop IoT networks.
    \item \textbf{Cross-Edge Continuity:} Develop distributed session ledgers that enable secure session migration across federated or mobile edge gateways.
    \item \textbf{Formal Security Verification:} Employ formal frameworks such as BAN logic, ProVerif, or Tamarin to validate confidentiality, authentication, and replay-resistance guarantees under active-adversary models.
    \item \textbf{Energy and Resource Profiling:} Perform long-term current and CPU utilization measurements across diverse MCU platforms (ESP32–S3, STM32, Nordic nRF) to quantify lifetime energy cost.
    \item \textbf{Integration with AI-Driven Analytics:} Incorporate adaptive machine-learning–based anomaly detection and trust evaluation within DSEKP-protected data streams for self-learning edge security.
\end{itemize}

Overall, DSEKP demonstrates that session-aware cryptographic agility can be realized even on low-power IoT hardware without compromising latency, throughput, or scalability. 
It represents a concrete and forward-looking step toward secure, autonomous, and intelligent edge computing infrastructures capable of sustaining the demands of future IoT ecosystems.


\section*{Acknowledgments}
\label{sec:acknowledgments}

The authors express their sincere gratitude to the \textbf{Department of Computer and System Sciences, Siksha-Bhavana, Visva-Bharati University, Santiniketan – 731235}, for providing a supportive research environment and the essential facilities required to carry out this work.  
Special thanks are also due to the \textbf{University Grants Commission (UGC)} for providing financial assistance through the \textit{National Eligibility Test (NET) – Junior Research Fellowship (JRF)} program under Ref.~No.~210510078094.  

\vspace{0.5em}
\section*{Data Availability}
The PSK baseline benchmark dataset \cite{rg5z-ge26-25} and the proposed DSEKP dataset \cite{3vw8-3a13-25} used in this study are publicly archived on IEEE DataPort. These resources include encrypted and decrypted MQTT Dataset logs to support reproducibility.

\newpage


\bibliographystyle{elsarticle-num}
\bibliography{refs}

\end{document}